\documentclass[aps,prd,onecolumn,showpacs,amsmath,amssymb,nofootinbib]{revtex4}
\usepackage{graphicx}
\usepackage{dcolumn}
\usepackage{bm}
\def\beqra{\begin{eqnarray}}
\def\eeqra{\end{eqnarray}}
\def\beq{\begin{equation}}
\def\eeq{\end{equation}}

\begin{document}
\newcommand{\Spin}[4]{\, {}_{#2}^{\vphantom{#4}} {#1}_{#3}^{#4}}
\newcommand{\vertsp}{\vphantom{\displaystyle{\dot a \over a}}}
\newcommand{\Gm}[3]{\, {}_{#1}^{\vphantom{#3}} G_{#2}^{#3}}
\newcommand{\Spy}[3]{\, {}_{#1}^{\vphantom{#3}} Y_{#2}^{#3}}
\draft

\title{B-mode polarization of the CMB from the second-order photon quadrupole}

\author{Nicola Bartolo}\email{nicola.bartolo@pd.infn.it}
\affiliation{Dipartimento di Fisica ``G.\ Galilei'', Universit\`{a} di Padova\\  
        and INFN -
Sezione di Padova \\ via Marzolo 8, I-35131 Padova, Italy}
\author{Sabino Matarrese}\email{sabino.matarrese@pd.infn.it}
\affiliation{Dipartimento di Fisica ``G.\ Galilei'',
Universit\`{a} di Padova\\  INFN -
Sezione di Padova \\ via Marzolo 8, I-35131 Padova, Italy}
\author{Silvia Mollerach}\email{mollerach@cab.cnea.gov.ar}
\affiliation{Centro At\'omico Bariloche, Av. Bustillo 9500 \\ 8400
Bariloche, Argentina}
\author{Antonio Riotto}\email{antonio.riotto@pd.infn.it}
\affiliation{D\'epartement de Physique Th\'eorique, Universit\'e de
Gen\`eve, 24 Quai Ansermet,
Gen\`eve, Switzerland,\\
and  INFN -
Sezione di Padova via Marzolo 8, I-35131 Padova, Italy}

\date{\today}
\begin{abstract}
We study a new contribution to the polarization of the Cosmic Microwave
Background induced at the epoch of recombination by the second-order 
quadrupole moment of the photon distribution. 
At second order in perturbation theory the quadrupole moment is not 
suppressed by the inverse of the optical depth in the tight coupling
limit, as it happens at first order in perturbation theory. We concentrate on  
the B-mode CMB polarization and find that such a novel contribution 
constitutes a contamination in the detection of the primordial tensor 
modes if the tensor to scalar ratio $r$ 
is smaller than a few $\times 10^{-5}$. The magnitude of the effect is 
larger than the B-mode due to secondary vector/tensor
perturbations and the analogous effect generated during the reionization
epoch, while it is smaller than the contamination produced by the conversion of polarization  of type E into type B,
by weak gravitational lensing. However the lensing signal can be 
cleaned, making the secondary modes discussed here the actual contamination 
limiting the detection of small amplitude primordial gravitational 
waves if $r$ is below $\simeq 10^{-5}$.

\end{abstract}
\pacs{98.70.Vc,98.80.Es,98.80.Cq  \hfill DFPD 07/A/05}
\maketitle

\section{Introduction}
The hunting for the polarization of the Cosmic Microwave Background (CMB) 
is one of the major targets of present and future 
planned experiments since it offers a unique opportunity to 
gather information about both the early (inflationary) universe and more 
recent epochs. The polarization is generated by the anisotropic scattering 
of CMB photons off free electrons. The 
decomposition into E and B-modes~\cite{se97,ka97} 
is extremely useful when dealing with the nature of the cosmological 
perturbations that 
are responsible for the temperature and polarization anisotropies of the 
CMB: the B-modes are excited only by vector and 
tensor perturbations, thus allowing eventually to disentangle their effect 
from that of the scalar perturbations. 
One of the fundamental predictions of inflationary models is the generation 
of a stochastic background of gravitational waves 
(tensor perturbations of the metric) which makes the search for the B-mode 
polarization crucially linked 
to a probe of the inflationary scenario. The amplitude of this background 
is determined by the energy scale of inflation, 
which can widely vary among different inflationary models. From this point of view, 
future satellite missions, such as {\em Planck}, will have enough 
sensitivity to either detect or constrain the B-mode CMB polarization 
predicted by the simplest inflationary models. 
A lot of experimental and theoretical issues are currently under 
investigations in order to assess the feasibility of a detection of the 
primordial gravity waves via the B-mode polarization. 
The main problematic aspects come from  foregrounds~\cite{AHS}, instrumental 
noise~\cite{instr}, and 
the gravitational lensing on the CMB by the matter distribution. 
These effects might actually mask the signal due to primordial tensor modes. 
As far as the lensing ``contamination'' is concerned, this is a non-linear effect 
that implies the transformation of E-mode into B-mode polarization~\cite{zalsel98} and it  
has been pointed out that the inflationary 
gravitational-wave background can only be detected by CMB polarization 
measurements if the tensor to scalar ratio $r \ge
10^{-4}$, which corresponds to an energy scale of inflation 
larger than $3\times 10^{15}$ GeV \cite{knox,kesden,kesden03,kinney}. 
However, a better technique to {\it clean} polarization maps
from the lensing effect has been proposed, which would allow tensor-to-scalar 
ratios as low as $10^{-6}$, or even smaller, to be probed 
\cite{hirata,seljak03}. 

In this paper we point out the existence of a new contribution to the 
B-mode polarization. 
It is due to the observation made in 
Ref.~\cite{PII} that at second-order in perturbation theory the quadrupole moment of the photons 
is not suppressed by the inverse of the optical depth in 
the tight coupling limit, as it happens at first order in perturbation theory. 
The anisotropic Thomson scattering of the photon 
quadrupole leads to a polarization of the CMB. The study of the 
temperature anisotropies at second-order have been analyzed 
in several works~\cite{secondT} and recently Refs.~\cite{PI,PII} 
studied the evolution of the temperature anisotropies in the tight 
coupling regime through the full derivation of the Boltzmann  
equations for the photon, baryon and cold dark matter 
components. In Ref.~\cite{MHM} it has been shown how B-mode polarization is produced 
during the last scattering and reionization epochs by the non-linear 
evolution of cosmological perturbations which generates unavoidably vector and tensor
modes, starting from the primordial (scalar) density fluctuations. 

In our case it is the non-suppression of the second-order 
quadrupole in the tight coupling regime 
that makes the contribution to the B-mode polarization considered
here of a different origin than that of Ref.~\cite{MHM}. 
Our findings show that the corresponding level of the B-mode is 
larger and also that it overcomes by two orders of 
magnitude the analogous signal generated during the reionization epoch~\cite{HuR}. 
It turns out that such a novel contribution to the CMB B-mode constitutes
a background to the detection of the primordial gravitational waves effect if the tensor to scalar ratio 
is smaller than a few $\times 10^{-5}$.  The magnitude of the effect is smaller than the
contamination produced by the conversion of polarization  of type E into type 
B, by weak gravitational lensing. However the lensing signal can be
cleaned, making the secondary modes discussed here the actual
background contaminating the detection of small amplitude primordial
gravitational waves.

These secondary effects always exist 
and their amplitude has a one-to-one relation with the level of 
density perturbations, which is severely constrained 
by both CMB anisotropy measurements and Large-Scale Structure observations. 
Therefore, their properties are largely inflation model-independent, 
contrary to primary tensor modes whose amplitude is not only 
model-dependent, but is well-known to be suppressed in some cases, like e.g. 
in the so-called curvaton model for the generation of 
curvature perturbations \cite{curvaton}.

The plan of the paper is as follows. In Sec.~\ref{I} we 
recall the origin of the quadrupole moment in the tight coupling 
limit at second-order in perturbation theory. In Sec.~\ref{II} the contribution to the 
power spectrum of the B-mode polarization from the second-order quadrupole is computed 
and finally in Sec.~\ref{III} we compare the results to the 
contribution to the B-mode from the inflationary tensor modes and 
the gravitational lensing and with other secondary effects as well.

\section{Second-order quadrupole moment of photons and the tight coupling limit}
\label{I}

In this section we recall the starting point of our paper: at second (and higher) order 
in the perturbations the quadrupole moment of the photon distribution cannot be neglected even 
in the limit of tight coupling. At linear order the quadrupole and higher order moments 
of the photons, in the tight coupling limit, are negligible  
with respect to the first two moments (the energy density and velocity) because they turn out 
to be suppressed by increasing powers of the inverse optical depth $\tau^{-1}$. However in the 
non-linear regime there is a source for the quadrupole made up of linear-velocity squared terms 
which do not vanish in the tight coupling limit. 
Such a result has been obtained in details in Ref.~\cite{PII}, and in the following we 
limit ourselves to give some definitions and summarize the main steps necessary 
to achieve it. 

The quadrupole moment of the photon distribution is defined as 
\begin{equation}
\label{quadrupole}
\Pi^{ij}_{\gamma}=\int\frac{d\Omega}{4\pi}\,\left(n^i n^j-\frac{1}{3}
\delta^{ij}\right) \Delta \, ,
\end{equation} 
where $\Delta=\Delta^{(1)}+\Delta^{(2)}/2$ are the photon temperature anisotropies 
split into a first and a second-order part. They are given by 
\begin{equation}
\label{Delta2}
\Delta(x^i,n^i,\eta)=\frac{\int dp p^3\,  \delta f}{\int dp p^3 f^{(0)}}\, ,
\end{equation}
with $\eta$ the conformal time, and $f^{(0)}$ the zeroth-order (Bose-Einstein) value  
around which the distribution function of photons is 
expanded, $f=f^{(0)}+\delta f=f^{(0)}+f^{(1)}+f^{(2)}/2$.
Eq.~(\ref{Delta2}) represents the photon fractional energy perturbation (in a given direction), 
which is the integral of the photon distribution function perturbation $\delta f$ 
over the photon momentum magnitude $p$ ($p^i=p n^i$). 
The angular dependence of the photon anisotropies $\Delta$ can be expanded as   
\begin{equation}
\label{Dlm2}
\Delta({\bf x}, {\bf n})=\sum_{\ell} \sum_{m=-\ell}^{\ell} \Delta_{\ell m}({\bf x})  
(-i)^{\ell}  \sqrt{\frac{4 \pi}{2\ell+1}}Y_{\ell m}({\bf n})\, ,
\end{equation} 
with  
\begin{equation}
\label{angular1}
\Delta_{\ell m}=(-i)^{- \ell}\sqrt{\frac{2\ell+1}{4\pi}} \int d\Omega  \Delta 
Y^{*}_{\ell m}({\bf n}) \, .
\end{equation}
The Boltzmann equations up to second-order for the photon temperature anisotropies $\Delta$ (
and the hierarchy equations for $\Delta_{\ell m}$) have been obtained in Ref.~\cite{PI}, together with the 
second-order Boltzmann equations for the baryons and cold dark matter. 
One can integrate the R.H.S. of the Boltzmann equation for $\Delta^{(2)}$  
over $d\Omega_{\bf n} (n^in^j-\delta^{ij}/3)/4\pi$ and set it to be vanishing in the limit of tight coupling. 
Then the  R.H.S. becomes~\cite{PII} 
\begin{equation}
\label{2quadRHS}
\frac{\dot{\tau}}{2} \left[ 
-\Pi^{ij}_\gamma+\frac{1}{10} \Pi^{ij}_\gamma+(\delta_b+\Psi) \left(-2 \Pi^{ij}_\gamma-\frac{1}{25} 
\Delta_{20} 
   (\hat{v}^i\hat{v}^j-\frac{1}{3} \delta^{ij}) \right)  +\frac{12}{5}\left(v^iv^j-\frac{1}{3} \delta^{ij} 
v^2  \right) 
\right]\, ,
\end{equation}
where $v^i$ is the photon-baryon velocity, $\delta_b$ the baryon density contrast, 
$\Psi$ the gravitational potential, and  
the differential optical depth $\dot \tau = {\bar n}_e \sigma_T a$ sets 
the collision rate in conformal time, with ${\bar n}_e$ the mean free electron density
and $\sigma_T$ the Thomson cross section.\footnote{The perturbed line element around a 
spatially flat FRW background is taken in the so-called Poisson 
gauge~\cite{Bert,mmb}, which eliminates one scalar degree of freedom from the $g_{0i}$ 
component and one scalar and two 
vector degrees of freedom from $g_{ij}$, thus reducing at linear order to the Newtonian gauge
\begin{equation}
\label{linelem}
d s^2 = a^2(\eta)\left\{-\left(1+2\Psi\right)d\eta^2 
- 2 V_i d\eta dx^i + \left[\left(1-2\Phi\right)\delta_{ij} + 
2 H_{ij}\right]dx^i dx^j \right\} \, .
\end{equation}
Each perturbation can be decomposed into a linear and a 
second-order term, e.g, $\Psi=\Psi^{(1)}+\Psi^{(2)}/2$.
$V_i$ contains scalar and vector (divergence-free) perturbation modes, while $H_{ij}$ is a 
tensor mode (i.e. traceless and divergence-free, $\partial^i H_{ij}=H^i_{~i}=0)$.} 
Therefore, in the limit of tight coupling, when the interaction rate is very high, the second-order 
quadrupole moment is given by 
\begin{equation}
\label{2quad}
\Pi^{ij}_\gamma \simeq \frac{8}{3} \left(v^iv^j-\frac{1}{3} \delta^{ij} v^2  \right)\, ,
\end{equation}
as it it follows by requiring that Eq.~(\ref{2quadRHS}) vanishes 
(the term multiplying $(\delta_b+\Psi)$ goes to zero in the tight 
coupling limit since it just comes from the first-order collision term). 
At linear order one would simply get the term $(- 9 \dot{\tau} \Pi^{ij}_\gamma/10)$ 
implying that, in the limit of a high scattering rate $\dot{\tau}$,
$\Pi^{ij}_\gamma$ goes to zero. However at second-order the quadrupole is not suppressed in the tight 
coupling limit because it turns out to be sourced by the linear-velocity squared. 

Indeed the result of Eq.~(\ref{2quad}) is not surprising and had to be expected. First, 
the limit of tight coupling corresponds to treat 
the photons in the fluid approximation, and, at second-order, the energy momentum tensor of 
a fluid is exactly characterized by  an anisotropic stress given by Eq.~(\ref{2quad}) (see, for example, Ref.~\cite{enh}) 
Moreover it is also known that, besides the usual dipole contribution, 
beyond linear order a quadratic Doppler effect gives rise to a quadrupole 
anisotropy (see Refs.~\cite{SuZe,Huthesis,HuR}).

\section{CMB Polarization}
\label{II}

Since the polarization  of the CMB is generated by the 
scattering of the quadrupole anisotropies by free electrons, armed with Eq.~(\ref{2quad}), 
we will explore the effects of such a contribution to the quadrupole on 
the CMB polarization.\footnote{The effects of the 
quadrupole moment~(\ref{2quad}) on the CMB temperature anisotropies have 
already been studied in Ref.~\cite{PII}.} Notice that the resulting polarization 
power spectrum has a different origin 
from that computed in Ref.~\cite{MHM}. Ref.~\cite{MHM} shows how 
second-order vector and tensor perturbations can produce CMB polarization through 
the decoupling phase. Once the vector and the tensor perturbations are obtained at 
second-order as a non-linear combination 
of first-order scalar perturbations, the induced power spectra of the E- and B-mode 
polarization are computed using the 
standard techniques developed at linear order~\cite{ZH}: 
a quadrupole is generated by the vector and tensor perturbations only beyond the 
leading order in the tight coupling expansion (otherwise it is suppressed) 
and to evaluate it one has to carefully account for the epoch around decoupling.
The result is thus proportional to 
$\Delta \eta_*$, the width of the visibility function (i.e, the thickness of the last scattering surface). 
The case of Eq.~(\ref{2quad}) 
is completely different in that 
it is a quadrupole that is indeed generated in the tight coupling limit, and the 
corresponding power spectra of 
CMB polarization will be computed following different approximations as well.                
          
\subsection{Polarization angular power spectra}

In order to study the generation of polarization anisotropies we will 
employ the formalism of the total 
angular momentum method~\cite{hw,hu98} 
which has the advantage of putting scalar, vector and tensor perturbations on an 
equal footing. This is the most convenient approach for our 
calculations, and we are going to extensively use the results of Ref. 
\cite{hw}. 
The temperature and polarization fluctuations are expanded in normal modes
that take into account the dependence on both the angular direction of 
photon propagation ${\bf n}$ and the spatial position ${\bf x}$,
$\Spin{G}{s}{\ell}{m}({\bf x},{\bf n})$ \footnote{In order to avoid confusion with 
the notations of Ref.~\cite{hw}, 
from now on, if not explicitly written, we are going to drop the superscript denoting 
the order of the perturbations.}

\begin{equation}
\begin{array}{rcl}
\Theta({\bf x},{\bf n},\eta) &=& \displaystyle{
\int \frac{d^3 k}{(2\pi)^3} }
        \sum_{\ell}
\sum_{m=-2}^2 \Theta_\ell^{(m)} \Gm{0}{\ell}{m} \, , \\
 (Q \pm i U)({\bf x},{\bf n},\eta) &=& \displaystyle{\int \frac{d^3k}
 {(2\pi)^3}}
        \sum_{\ell} \sum_{m=-2}^2
        (E_\ell^{(m)} \pm i B_\ell^{(m)}) \, \Gm{\pm 2}{\ell}{m} \,, 
\end{array}
\label{eqn:decomposition}
\end{equation}
with spin $s=0$ describing the temperature fluctuation and $s=\pm 2$
describing the polarization tensor and $m=0, \pm 1, \pm 2$ denoting scalar, 
vector and tensor perturbations, respectively.
$E_\ell^{(m)}$ and $B_\ell^{(m)}$ are the angular moments of the
electric and magnetic polarization components and  
\begin{equation}
\Gm{s}{\ell}{m}({\bf x},{\bf n})  =
        (-i)^\ell \sqrt{ \frac{4\pi}{2\ell+1}}
        [\Spy{s}{\ell}{m}({\bf n})] \exp(i{\bf k} \cdot {\bf x})\,.
\end{equation}

Before keeping on recalling the main results of the total angular 
momentum method, a comment is in order here. 
The Boltzmann equations for the temperature and polarization anisotropies we are going to write 
are derived within linear perturbation theory. Then, given the 
second-order quadrupole~(\ref{2quad}), our approach will 
be that of evolving it linearly into Eqs.~(\ref{eqn:pol}). In fact this is 
the correct procedure if we wish to isolate its contribution 
to the CMB polarization at second-order, distinguishing it from other 
unavoidable corrections to the equations~(\ref{eqn:pol})
coming in the form of first-order squared terms. In a similar way, the 
variable $\Delta^{(2)}({\bf x}, {\bf n}, \eta)$ 
will be taken to be $4 \Theta({\bf x}, {\bf n}, \eta)$, a relation holding 
at the linear level (see Refs.~\cite{JD,PII}).

The Boltzmann equation describing the time evolution of the radiation 
distribution under gravitation and scattering processes can be written as a
set of evolution equations for the angular moments of the temperature,
$\Theta_\ell^{(m)}$ (for $\ell \ge m)$, and both polarization types, 
$E_\ell^{(m)}$ and $B_\ell^{(m)}$ (for $\ell \ge 2$ and $m \ge 0$), 
\begin{eqnarray}
\dot \Theta_\ell^{(m)}&=& k
\Bigg[ \frac{\Spin{\kappa}{0}{\ell}{m}}{(2\ell-1)}
        \Theta_{\ell-1}^{(m)}
             -\frac{\Spin{\kappa}{0}{\ell+1}{m}}{(2\ell+3)}
        \Theta_{\ell+1}^{(m)} \Bigg]
	- \dot\tau \Theta_\ell^{(m)} + S_\ell^{(m)}, \\
\label{eqn:boltzmannE}
\dot E_\ell^{(m)}&=& k \Bigg[ {\Spin{\kappa}{2}{\ell}{m} \over (2\ell-1)}
E_{\ell-1}^{(m)} - {2m \over \ell (\ell + 1)} B_\ell^{(m)} 
- {\Spin{\kappa}{2}{\ell+1}{m} \over (2 \ell + 3)}
        E_{\ell + 1}^{(m)} \Bigg] 
	- \dot\tau [E_\ell^{(m)}+\sqrt{6}P^{(m)}\delta_{\ell,2}]\, , \\
\dot B_\ell^{(m)}&=& k \Bigg[ {\Spin{\kappa}{2}{\ell}{m} \over (2\ell-1)}
B_{\ell-1}^{(m)} + {2m \over \ell (\ell + 1)} E_\ell^{(m)} 
- {\Spin{\kappa}{2}{\ell+1}{m} \over (2 \ell + 3)}
        B_{\ell + 1}^{(m)} \Bigg] -\dot\tau B_\ell^{(m)},
\label{eqn:boltzmann}
\end{eqnarray}
where  a dot stands for a derivative with respect to the conformal time $\eta$ and the coupling coefficients are
\begin{equation}
\Spin{\kappa}{s}{\ell}{m} = \sqrt{ 
{(\ell^2-m^2)(\ell^2-s^2)\over\ell^2}}.
\end{equation}
The fluctuation sources are given by
\begin{equation}
\begin{array}{lll}
S_0^{(0)} = \dot\tau \Theta_0^{(0)} -  \dot\Phi \, ,      \qquad &
S_1^{(0)} = \dot\tau v_B^{(0)} + k\Psi \, ,               \qquad &
S_2^{(0)} = \dot\tau P^{(0)} \, , \vertsp\\
                                                \qquad &
S_1^{(1)} = \dot\tau v_B^{(1)} + \dot V \, ,              \qquad &
S_2^{(1)} = \dot\tau P^{(1)} \, , \vertsp\\
                                                \qquad &
                                                \qquad &
S_2^{(2)} = \dot\tau P^{(2)} - \dot H      \vertsp \, ,
\label{eqn:S}
\end{array}
\end{equation}
with
\begin{equation}
P^{(m)} = {1 \over 10} \left[ \Theta_2^{(m)}  -
{\sqrt 6} E_2^{(m)} \right]. 
\label{eqn:polsource}
\end{equation}
The modes with $m=-|m|$ satisfy the same equations with  $B_\ell^{(-|m|)}= 
-B_\ell^{(|m|)}$ and all the other quantities unchanged. In Eq.~(\ref{eqn:S}) $v_B$ is the baryon velocity perturbation which in 
the tight coupling regime corresponds to the photon-baryon velocity $v$. 

These equations can be formally integrated, leading to  
simple expressions in terms of an integral along the line-of-sight \cite{hw}. 
The temperature fluctuations are given by
\begin{equation}
{\Theta_\ell^{(m)}(\eta,k) \over 2\ell + 1}\, 
  =  
\int_0^{\eta} d\eta' \, e^{-\tau} \sum_{\ell'} \,
  S_{\ell'}^{(m)}(\eta') \, j_\ell^{(\ell'm)}(k(\eta-\eta')) \, ,
\label{eqn:inttemp}
\end{equation}
where $j_\ell^{(\ell'm)}$ are given in Ref.~\cite{hw}.
For the polarization, we have
\begin{eqnarray}
{E^{(m)}_\ell(\eta_0,k) \over 2\ell+1} &=&  -\sqrt{6}
\int_0^{\eta_0} d\eta \, \dot\tau e^{-\tau} P^{(m)}_{\vphantom{\ell}}
(\eta)\epsilon_\ell^{(m)}(k(\eta_0-\eta)) \, ,\nonumber\\
{B^{(m)}_\ell(\eta_0,k) \over 2\ell+1} &=&  -\sqrt{6} 
\int_0^{\eta_0} d\eta \, \dot\tau e^{-\tau} P^{(m)}_{\vphantom{\ell}}
(\eta)\beta_\ell^{(m)}(k(\eta_0-\eta)) \, ,
\label{eqn:pol}
\end{eqnarray}
where the radial functions read
\begin{eqnarray}
\epsilon^{(\pm 1)}_\ell(x) &=& {1 \over 2} {\sqrt {(\ell-1)(\ell+2)}}
        \left[ {j_\ell(x) \over  x^2} + {j_\ell'(x) \over  x}
        \right] \, , \nonumber\\ 
\epsilon^{(\pm 2)}_\ell(x) &=& {1 \over 4} \left[ -j_\ell(x)
        + j_\ell''(x) + 2{j_\ell(x) \over x^2} +
        4{j_\ell'(x) \over x} \right]  \, , \\
\beta^{(+1)}_\ell(x) & = & - \beta^{(-1)}_\ell(x) = 
{1 \over 2} {\sqrt {(\ell-1)(\ell+2)}}~
        {j_\ell(x) \over x} \, , \nonumber\\
\beta^{(+2)}_\ell(x) &=& -\beta^{(-2)}_\ell(x) = 
{1 \over 2} \left[ j_\ell'(x)
        + 2 {j_\ell(x) \over x} \right] \, .
\end{eqnarray}
The differential optical depth $\dot \tau = {\bar n}_e \sigma_T a$ sets 
the collision rate in conformal time, with ${\bar n}_e$ the mean free electron density
and $\sigma_T$ the Thomson cross section and
$\tau(\eta_0,\eta) \equiv \int_\eta^{\eta_0} \dot\tau(\eta') 
d\eta^\prime$ 
the optical depth between $\eta$ and the present time. 
The combination $g(\eta)=\dot\tau e^{-\tau}$ is the {\em visibility function}
and it expresses the probability that a photon last scattered between
$d\eta$ of $\eta$ (with $\int d\eta g(\eta)=1$) and hence it 
is sharply peaked at the last scattering epoch. In early reionization models, a second peak is also present at
more recent times.  

Scalar modes do not contribute to B-polarization, thus
$B^{(0)}_\ell=0$. 
We are interested in the contribution to the angular power-spectrum 
for the E and B modes arising from vector ($m=1$) and tensor ($m=2$)
perturbations, 
\begin{eqnarray}
C_\ell^{(E)\hphantom{E}} 
		  & = &{2 \over \pi} \int {dk \over k}
	\sum_{m=-2}^2
        k^3 \frac{|E_\ell^{(m)}(\eta_0,k)|^2}{(2\ell+1)^2} , \nonumber\\
\label{CLB}
C_\ell^{(B)\hphantom{E}} 
		  & = &{2 \over \pi} \int {dk \over k}
	\sum_{m=-2}^2
        k^3 \frac{|B_\ell^{(m)}(\eta_0,k)|^2}{(2\ell+1)^2}. 
\end{eqnarray}

In the following we will focus only 
on the $B$-mode of CMB polarization. This is probably the most 
interesting case when confronting the signal generated by the primordial 
gravitational waves with that coming from second-order perturbations. 
Both are expected to be very low, but for different reasons: 
the former because of the low content of primordial gravitational waves, 
the latter because it is a second-order effect. 
Then, as first pointed out in Ref.~\cite{MHM}, there could be the 
possibility that for a low value of primordial tensor to 
scalar ratio $r$, this kind of secondary signal could constitute an 
additional barrier to our ability in detecting the signature 
of the inflationary gravitational waves.  

\subsection{B-mode polarization from the tight coupling quadrupole}

The first step is to compute the polarization source term $P^{(m)}$ in Eq.~(\ref{eqn:pol}). 
In the tight coupling limit we deduce from 
Eq.~(\ref{eqn:boltzmannE}) that $\Theta^{(m)}_2=-4 E^{(m)}_2/\sqrt{6}$, 
where we have used Eq.~(\ref{eqn:polsource}). 
Plugging this result back into Eq.~(\ref{eqn:polsource}) yields 
\begin{equation}
\label{Pmtight}
P^{(m)}=\frac{\Theta_2^{(m)}}{4} \, .
\end{equation}
The B-mode polarization field is then obtained from Eq.~(\ref{eqn:pol}) 
exploiting the fact that the visibility function is sharply peaked 
around the epoch of decoupling $\eta_*$. We thus evaluate $P^{(m)}(\eta)$ and  
$\beta_\ell^{(1)}(k(\eta_0-\eta))$ at $\eta_*$ taking them 
out of the integrals
\begin{equation}
\label{Btight}
\frac{B^{(m)}_\ell(k,\eta_0)}{2\ell+1}=\frac{\sqrt{6}}{16} \Delta_{2m}(k,\eta_*) 
\beta_\ell^{(1)}(k(\eta_0-\eta_*))\, , 
\end{equation}   
where we have used $\Delta({\bf x}, {\bf n}, \eta)=4 \Theta({\bf x}, {\bf n}, \eta)$. 
Notice the different approximations used in deriving Eq.~(\ref{Btight}) with 
respect to the procedure of Ref.~\cite{MHM}; in particular 
we do not need to go beyond the leading order in the tight coupling 
approximation, and this allows an easier evaluation of the time integral~(\ref{eqn:pol}).  

The quantities $\Delta_{2m}$ are related to the quadrupole defined in Eq.~(\ref{quadrupole}). 
If in Fourier space for a given ${\bf k}$ we 
choose ${\bf e}_3={\hat{\bf k}}$ and ${\bf e}_1$ and ${\bf e}_2$ forming an orthonormal basis, 
it is easy to verify that 
\begin{eqnarray}       
\Delta_{20}&=&-\frac{15}{2} e_{3i}\,e_{3j}\, \Pi^{ij}_\gamma\, , \\
\label{D21}
\Delta_{2\pm 1}&=&\pm 5 \sqrt{3} \, e_{3i}\,  \frac{({\bf e}_1\mp i {\bf e}_2)_j}{\sqrt{2}} \, 
\Pi^{ij}_\gamma \, , \\
\label{D22}
\Delta_{2\pm 2}&=&-5 \sqrt{\frac{3}{2}} \frac{({\bf e}_1\mp i {\bf e}_2)_i}{\sqrt{2}}\,   
\frac{({\bf e}_1\mp i {\bf e}_2)_j}{\sqrt{2}}
 \, \Pi^{ij}_\gamma \, ,
\end{eqnarray}
where one has to use Eq.(\ref{angular1}) and the explicit expression of the spherical 
harmonics in the chosen coordinate system. 
Being interested in the B-mode polarization in the following we will make use only of 
the last two expressions, for the 
multipoles $\pm 1$ and $\pm 2$, corresponding to the vector and tensor parts of the quadrupole moment.

The (firs-order) velocities appearing in~(\ref{2quad}) are those of the tightly coupled 
fluid of photons and baryons, 
and we will use the 
analytical solutions found in Ref.~\cite{PII} with  
\begin{equation}
\label{v1sol}
v^{i}_\gamma=-i \frac{k^i}{k}\frac{9}{10} \Phi^{(1)}_{\bf k}(0) \sin(k c_s \eta) c_s 
\quad (k \ll k_{eq})\, ,
\end{equation} 
 and 
\begin{equation}
\label{v1solbis}
v^{i}_\gamma=-i \frac{k^i}{k}\frac{9}{2} \Phi^{(1)}_{\bf k}(0) \sin(k c_s \eta) c_s 
\quad (k \gg k_{eq})\, ,
\end{equation} 
for modes entering the horizon before or after the equality epoch, respectively. 
Here $\Phi^{(1)}_{\bf k}(0)$ is the primordial linear 
gravitational potential on superhorizon scales deep in the radiation dominated 
epoch, while  $c_s=(3(1+R_*))^{-1/2}$ is the sound speed of the 
photon-baryon fluid, with $R_*=3\rho_b/(4\rho_\gamma)$ the ratio of the baryon 
to photon energy density at the decoupling epoch. 

Using Eq.~(\ref{v1sol}) and~(\ref{v1solbis}) in Eqs.~(\ref{D21}) and~(\ref{D22}) 
the temperature quadrupole anisotropies can be written as 
a convolution 
\begin{equation}
\label{gen}
\Delta_{2m}({\bf k}, \eta_*) =\int \frac{d^3k_1}{(2 \pi)^3} 
f_m({\bf k}_1,{\bf k}-{\bf k}_1) \Phi^{(1)}_{{\bf k}_1}(0) 
\Phi^{(1)}_{{\bf k}-{\bf k}_1}(0)\, ,
\end{equation}   
where the kernel for the vector part, $m= \pm 1$, reads 
\begin{eqnarray}
\label{K1}
f_{\pm 1}({\bf k}_1,{\bf k}-{\bf k}_1)
\hat {\bf k}_1^j \hat {\bf k}_2^i) 
&=& \mp \frac{20}{\sqrt{6}} N c_s^2 \sin(k_1c_s\eta_*) \sin(|{\bf k}-{\bf k}_1| 
c_s \eta_*)\,  e^{\mp i \phi_{{\bf k}_1}} 
\sin\theta_{{\bf k}_1} \Bigg[ \frac{k}{|{\bf k}-{\bf k}_1|} -2\cos\theta_{{\bf k}_1} 
\frac{k_1}{|{\bf k}-{\bf k}_1|} \Bigg]\, , 
\end{eqnarray}
a vector normalized to unity, e.g. 
and that for the tensor part, $m= \pm 2$, is given by 
\begin{eqnarray}
\label{K2}
f_{\pm 2}({\bf k}_1,{\bf k}-{\bf k}_1)&=& - \frac{20}{\sqrt{6}} N c_s^2 
\sin(k_1c_s\eta_*) \sin(|{\bf k}-{\bf k}_1| c_s \eta_*)\,  
e^{\mp 2 i \phi_{{\bf k}_1}} \sin\theta_{{\bf k}_1}  
\frac{k_1}{|{\bf k}-{\bf k}_1|}\, .   
\end{eqnarray}
To compute Eqs.~(\ref{K1}) and~(\ref{K2}) one makes use of the 
orthonormality of the basis ${\bf e}_3 =\hat{ \bf k}$, ${\bf e}_1$ and 
${\bf e}_2$; $\phi_{{\bf k}_1}$ and $ \theta_{{\bf k}_1}$ are the 
azimuthal and polar angles of ${\bf k}_1$ in such a coordinate system. 
The coefficient $N$ comes from the prefactors appearing in the expressions 
for the linear velocities~(\ref{v1sol}) and~(\ref{v1solbis}), with 
$N=(9/10)^2$ if both $k_1$ and $|{\bf k}-{\bf k}_1|$ are less that $k_{eq}$, 
$N=(9/2)^2$ if both the wavenumbers are greater than $k_{eq}$, 
while in the mixed cases $N=(9/10) \times  (9/2)$. 

The power spectrum of the quadrupole anisotropies is defined by 
\begin{equation}
\langle \Delta_{2m}({\bf k}) \Delta^*_{2m}({\bf p}) \rangle 
= (2 \pi)^3 P_{\Delta_{2m}}(k)\,  \delta^{(3)}({\bf k} -{\bf p})\, ,
\end{equation}
 and, using the general expression~(\ref{gen}) it turns out to be 
\begin{equation}
\label{PSgen}
P_{\Delta_{2m}}(k)=2 \int \frac{d^3k_1}{(2 \pi)^3} 
|f_m({\bf k}_1,{\bf k}-{\bf k}_1)|^2 P_{\Phi}(k_1) P_\Phi(|{\bf k}-{\bf k}_1)|)\, ,   
\end{equation}       
where $P_{\Phi}(k_1)$ is the power spectrum of the primordial gravitational 
potential.  The power spectrum of the quadrupole anisotropies 
enters in the quantity we are interested in, namely the contribution to the 
angular power spectrum of the B-mode from tensor and vector 
perturbations as given by Eq.~(\ref{CLB}). Inserting Eq.~(\ref{Btight}) we find 
\begin{eqnarray}
C^{(B)(m)}_\ell &=& \frac{2}{\pi} \int dk\, k^2 \frac{|B^{(m)}_\ell(k,\eta_0)|^2}{(2\ell+1)^2}
=\frac{3}{64 \pi} \int dk\, k^2 P_{\Delta_{2m}}(k,\eta_*) \,
\Big[\beta^{(m)}_\ell(k(\eta_0-\eta_*))    \Big]^2\, . \nonumber \\
\end{eqnarray}   

Let us now write down explicitly the power spectra~(\ref{PSgen}) using 
Eqs.~(\ref{K1})-({\ref{K2}). For the vector perturbation mode we find 
\begin{eqnarray}
\label{PDpm1}
P_{\Delta_{2\pm 1}}(k,\eta_*)&=&\frac{100}{3}  \frac{c_s^4}{\pi^2} 
\int_{-1}^{+1} d(\cos \theta_{{\bf k}_1}) (1-\cos^2 \theta_{{\bf k}_1}) 
\int_0^{\infty}dk_1\,  N^2 k_1^2\, \sin^2(k_1 c_s \eta_*) \sin^2(|{\bf k}-{\bf k}_1| c_s \eta_*) 
\nonumber \\
&\times& \Bigg[ \frac{k}{|{\bf k}-{\bf k}_1|} -2\cos\theta_{{\bf k}_1} 
\frac{k_1}{|{\bf k}-{\bf k}_1|} \Bigg]^2 P_\Phi(k_1) 
P_\Phi\left(\sqrt{k^2+k_1^2-2kk_1\cos \theta_{{\bf k}_1}}\right)\, ,
\end{eqnarray}
while for the tensor mode 
\begin{eqnarray}
\label{PDpm2} 
P_{\Delta_{2\pm 2}}(k,\eta_*)&=&\frac{100}{3}  \frac{c_s^4}{\pi^2} 
\int_{-1}^{+1} d(\cos \theta_{{\bf k}_1}) (1-\cos^2 \theta_{{\bf k}_1}) 
\int_0^{\infty}dk_1\,  N^2 \frac{k_1^4}{|{\bf k}-{\bf k}_1|^2} \, 
\sin^2(k_1 c_s \eta_*) \sin^2(|{\bf k}-{\bf k}_1| c_s \eta_*) 
\nonumber \\
&\times&  P_\Phi(k_1) P_\Phi\left(\sqrt{k^2+k_1^2-2kk_1\cos \theta_{{\bf k}_1}}\right)\, . 
\end{eqnarray}

An efficient way to handle the above integrals over the wavenumber 
${\bf k}_1$ is to introduce the variables
\begin{equation}
x=\frac{|{\bf k}-{\bf k}_1|}{k}\, , \quad \quad y=\frac{k_1}{k}\, .
\end{equation}
With such a substitution we arrive at a compact expression for the 
angular power spectra of the B-mode polarization
\begin{eqnarray}
\label{CBpm1}
C^{(B)(\pm 1)}_\ell&=&\frac{25}{144} \frac{1}{\pi^3} \int dk\, k^5 
\int_0^{\infty} dy\, y \int_{|1-y|}^{|1+y|} \frac{dx}{x} \Bigg[  1-\left(\frac{1+y^2-x^2}{2y} \right)^2
\Bigg] \nonumber \\
&\times&  (y^2-x^2)^2 N^2 \sin^2(kc_s\eta_*y) \sin^2(kc_s\eta_*x) P_\Phi(ky) P_\Phi(kx)  
\Big[\beta^{(\pm 1)}_\ell(k(\eta_0-\eta_*))    \Big]^2 \, , \nonumber \\
\end{eqnarray}
\begin{eqnarray}
\label{CBpm2}
C^{(B)(\pm 2)}_\ell&=&\frac{25}{144} \frac{1}{\pi^3} \int dk\, k^5 
\int_0^{\infty} dy\, y^3 \int_{|1-y|}^{|1+y|} \frac{dx}{x} 
\Bigg[  1-\left(\frac{1+y^2-x^2}{2y} \right)^2
\Bigg]^2  \nonumber \\
&\times& N^2 \sin^2(kc_s\eta_*y) \sin^2(kc_s\eta_*x) P_\Phi(ky) P_\Phi(kx)  
\Big[\beta^{(\pm 2)}_\ell(k(\eta_0-\eta_*))    \Big]^2 \, . \nonumber \\
\end{eqnarray}
We remind the reader that $N=(9/10)^2$ if both $k_1=k y$ and 
$|{\bf k}-{\bf k}_1|=k x $ are less that $k_{eq}$, $N=(9/2)^2$ if both the 
wavenumbers are greater than $k_{eq}$, while in the mixed cases 
$N=(9/10) \times  (9/2)$. The power spectrum of the primordial gravitational 
potential can be written as $P_\Phi=P_{0\Phi} k^{-3}(k/k_0)^{n_s-1}$, where 
$n_s$ is the scalar spectral index and the amplitude 
is related to that of the comoving curvature perturbation spectrum,
 $\Delta^2_{\cal R}(k) = \Delta^2_{\cal R}(k_0) 
(k/k_0)^{n_s-1}$, by $P_{0\Phi}=(4/9) 2\pi^2 \Delta^2_{\cal R}(k_0)$. 
Therefore Eqs.~(\ref{CBpm1}) and~(\ref{CBpm2}) become 
 \begin{eqnarray}
C^{(B)(\pm 1)}_\ell&=&P^2_{0\Phi} \frac{25}{144} \frac{1}{\pi^3} 
\int \frac{dk}{k} \, \left( \frac{k}{k_0}
\right)^{2(n_s-1)} 
\int_0^{\infty} dy\, y^{n_s-3} \int_{|1-y|}^{|1+y|} dx\, x^{n_s-5}  (y^2-x^2)^2 \nonumber \\
&\times&  \Bigg[  1-\left(\frac{1+y^2-x^2}{2y} \right)^2
\Bigg] N^2 \sin^2(kc_s\eta_*y) \sin^2(kc_s\eta_*x)   
\Big[\beta^{(\pm 1)}_\ell(k(\eta_0-\eta_*))    \Big]^2 \, , \nonumber \\
\end{eqnarray}
and
\begin{eqnarray}
C^{(B)(\pm 2)}_\ell&=&P^2_{0\Phi}  \frac{25}{144} \frac{1}{\pi^3} \int \frac{dk}{k} \, \left( \frac{k}{k_0}
\right)^{2(n_s-1)} 
\int_0^{\infty} dy\, y^{n_s-1} \int_{|1-y|}^{|1+y|} dx\, x^{n_s-5}  \nonumber \\
&\times&  \Bigg[  1-\left(\frac{1+y^2-x^2}{2y} \right)^2
\Bigg]^2 N^2 \sin^2(kc_s\eta_*y) \sin^2(kc_s\eta_*x)   
\Big[\beta^{(\pm 2)}_\ell(k(\eta_0-\eta_*))    \Big]^2 \, . \nonumber \\
\end{eqnarray}

To perform the integration in $k$ it is convenient to use the WKB
approximated expressions
\begin{eqnarray}
\langle (\beta_\ell^{(2)})^2(x) \rangle
& \simeq & \frac{1}{8} 
\left(\frac{{\sqrt {x^2-\ell^2}}}{x^3}+
\frac{4}{x^3{\sqrt {x^2-\ell^2}}}\right)\\
\langle (\epsilon_\ell^{(2)})^2(x) \rangle
& \simeq & \frac{1}{8} 
\left[\left(1-\frac{1+\ell(\ell+1)/2}{x^2}\right)^2
\frac{1}{x{\sqrt {x^2-\ell^2}}}+\frac{{\sqrt {x^2-\ell^2}}}{x^5}\right],
\nonumber
\label{be2}
\end{eqnarray}
for $x>\ell$, and 0 for $x<\ell$.

Notice that up to now we have not accounted for the diffusion 
damping effects in writing these integrals. This can be done by 
replacing the sines with
\begin{eqnarray}
\sin^2(kc_s\eta_*x) &\rightarrow& \sin^2(kc_s\eta_*x) e^{-2\frac{k^2x^2}{k^2_D}}\, , \\ 
\sin^2(kc_s\eta_*y) &\rightarrow& \sin^2(kc_s\eta_*y) e^{-2\frac{k^2y^2}{k^2_D}}\, .
\end{eqnarray} 
where $k_D$ is evaluated at the decoupling epoch and defines the damping scale~\cite{HSan}.

\section{Discussion and conclusions}
\label{III}

We show in Figure 1 the results for the vector and tensor contribution to the B-mode
polarization multipoles for a  $\Lambda$CDM model with scalar spectral 
index $n=0.95$ and normalization $\Delta^2_{\cal R}(k_0) = 2.3\times10^{-9}$
at $k_0$ = 0.002/Mpc (the damping scale has been chosen so that $k_D \eta_0 \simeq 1600$).
The tensor modes provide the dominant 
contribution for $\ell < 100$, while the vector modes contribution 
dominates at larger values of $\ell$. Notice that the structure of the 
peaks of the second-order effect is richer 
than the usual (linear) polarization power spectra 
because the quadrupole signal is modulated as a velocity squared 
giving rise to a $\sin^2(kc_s \eta_*)$ dependence, 
which appears with two different periods in the convolution products~(\ref{PDpm1})-(\ref{PDpm2}).   

\begin{figure}[t]
\includegraphics[width=17cm]{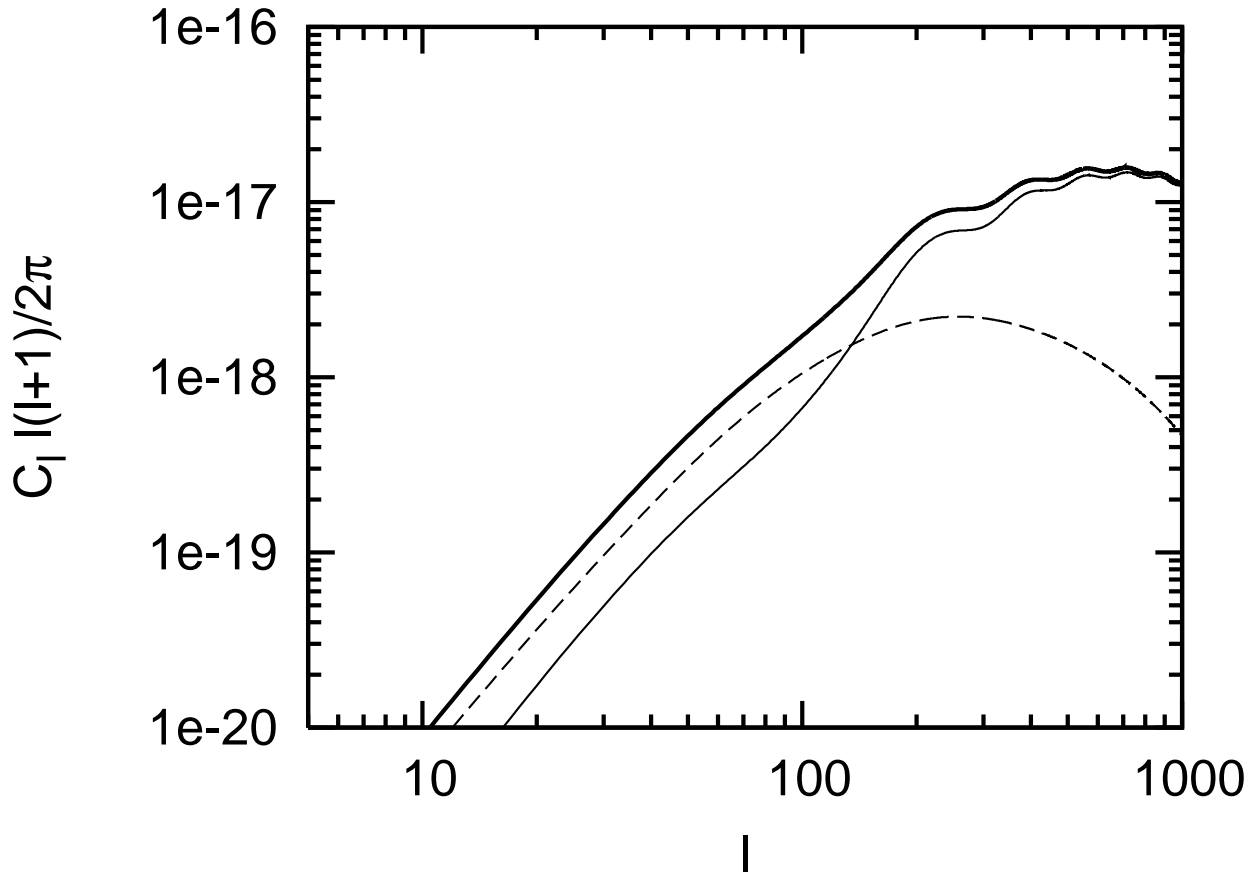}\\
\caption[fig0]{\label{fig1} Angular power-spectrum of 
B polarization originated at recombination by the second-order quadrupole. The thin solid line 
represents the  vector modes contribution and the dashed line the 
tensor modes. The thick solid line correspond to the total contribution.}

\end{figure}

Fig.~\ref{fig1} reveals that the second-order quadrupole B-mode polarization computed here is larger for 
$\ell >10$ than that due to the second-order vector and tensor modes, which are produced 
by the non-linear evolution of primordial scalar perturbations, considered in Ref.~\cite{MHM}. 
One could also ask what is the analogous signature on the B-mode CMB polarization induced by the 
second-order quadrupole moment of the photons during the reionization epoch. 
In this case the magnitude of the effect has been 
already computed in Ref.~\cite{HuR}. A comparison of our results 
in Fig.~\ref{fig1} to those shown in Fig.~9 
of Ref.~\cite{HuR} shows that the contribution of the second-order quadrupole from recombination 
overcomes the analogous signal 
generated during the reionization epoch by two orders of 
magnitude in the relevant range $100 \leq \ell \leq 1000$.    

It is also interesting to compare this second-order quadrupole contribution to the 
B-mode polarization with that expected from the primordial gravitational waves 
background generated during inflation. This depends of course on the model 
of inflation considered and its amplitude is parametrized by the ratio of the tensor ($h$) to 
scalar power spectrum amplitudes $r=\Delta^2_h(k_0)/\Delta^2_{\cal R}(k_0)$ (accordingly 
to the definition used in the WMAP analysis~\cite{wmap1infl,wmap3}). 
In Figure 2 we show the primordial gravitational wave B 
polarization multipoles for a model with $r = 10^{-5}$ (computed with the help 
of the CMBFAST code \cite{se96}), along with the total (vector plus tensor) 
contribution from the second-order quadrupole at recombination. 
We have considered for the $\Lambda$CDM model the cosmological parameters 
$\Omega_\Lambda=0.732$, $\Omega_m=0.268$, $\Omega_bh^2=0.0218$, $h=0.7$,
$\tau = 0.073$ \cite{wmap3}.
The contribution from the primordial 
gravitational waves scales linearly with $r$, thus we see that for values 
of $r$ smaller than a few $\times 10^{-5}$ the second-order quadrupole 
contribution becomes dominant for $\ell > 10$, limiting our ability to 
detect the primordial gravitational-wave background.
The peak from the primordial gravitational waves at the lowest multipoles, 
arising from the reionization epoch, could be affected by the contribution 
studied here, only for extremely low values of the scalar to tensor ratio 
($r < 10^{-8}$). 

The main background for the detection of the primordial gravitational waves 
through their B-mode polarization signal is however due to the weak gravitational
lensing conversion of the dominant E modes polarization to B modes 
\cite{zalsel98}.
Fortunately, this effect can be, at least partially, accounted for through 
reconstruction of the gravitational lensing potential by the correlations of
B-polarization between large and small angular scales, which primordial 
gravitational waves do not produce. ``Cleaning'' of the gravitational lensing 
signature may achieve factors of 40 in the power-spectrum, or even larger 
\cite{seljak03}.
In Figure 2 we display the predicted gravitational lensing signal in 
B polarization calculated with CMBFAST for the cosmological model under 
consideration, reduced by a factor of 40 \cite{hirata,seljak03}. 
Clearly, an improvement in the 
cleaning algorithm would make the second-order quadrupole polarization
derived here the actual limitation to the detection of
primordial gravitational waves through B-polarization of the CMB if 
$r <$ few $\times 10^{-5}$.

\begin{figure}[t]
\includegraphics[width=17cm]{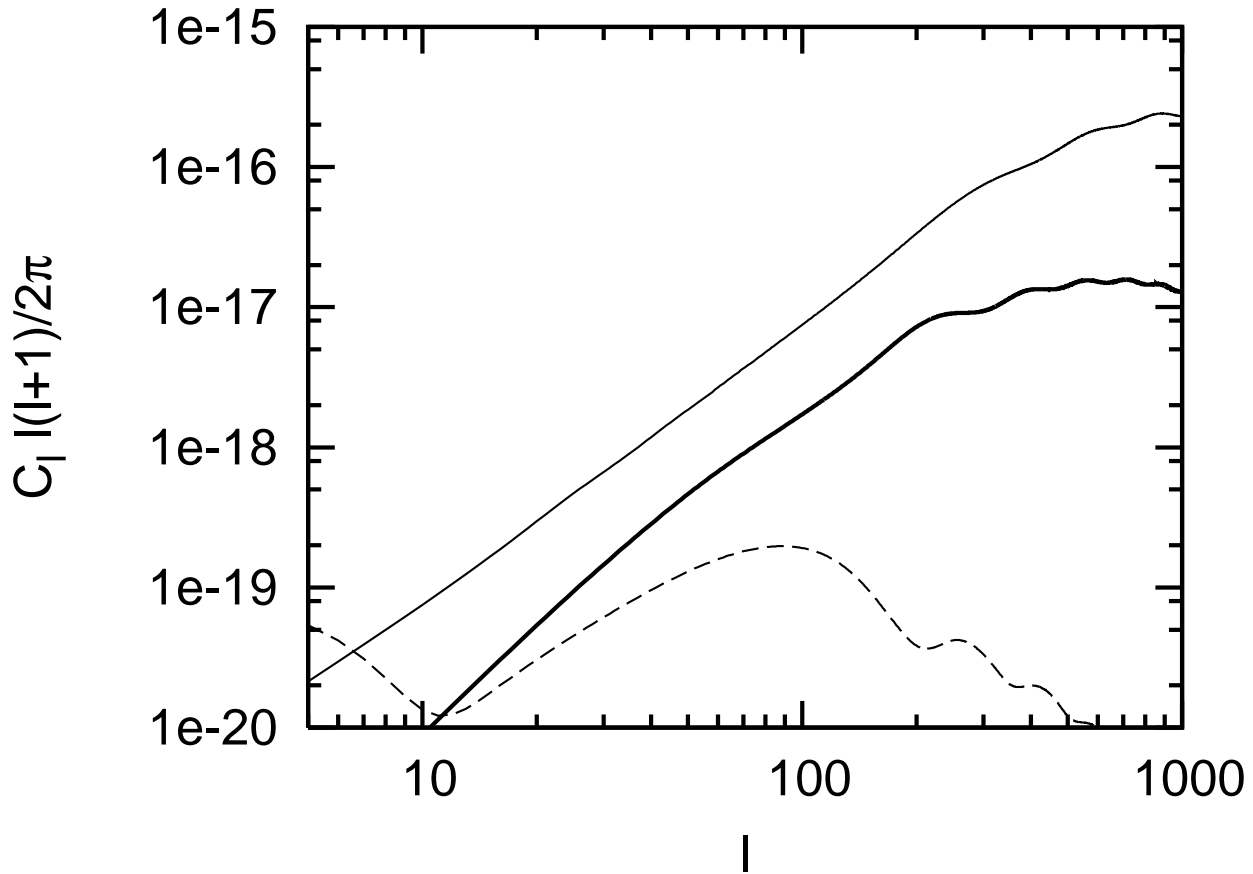}\\
\caption[fig0]{\label{fig2} Angular power spectrum of the total (vector plus 
tensor) B polarization generated at recombination by the second-order quadrupole 
(thick solid line) compared with that induced by primordial 
gravitational waves (dashed line)
with tensor to scalar ratio $r = 10^{-5}$ in a flat $\Lambda$CDM model. 
The thin solid line is the signal due to  gravitational lensing 
cleaned by a factor 40. }

\end{figure}

\acknowledgments{} 
We wish to thank Wayne Hu for e-mail exchanges. 


\end{document}